\documentclass[10pt,a4paper,final]{iopart}

\usepackage{iopams}  
\usepackage{graphicx}
\usepackage[breaklinks=true,colorlinks=true,linkcolor=blue,urlcolor=blue,citecolor=blue]{hyperref}

\begin{document}

\title[Comment]{Comment on Limitations on the superposition principle: superselection rules in non-relativistic quantum mechanics}

\author{Namit Anand$^{1}$}
\address{$^1$National Institute of Science Education and Research, Bhubaneswar, India}
\ead{namit.anand@niser.ac.in}

\begin{abstract}
This is a comment to the paper, Limitations on the superposition principle: superselection rules in non-relativistic quantum mechanics by C Cisneros et al 1998 Eur. J. Phys. 19 237. doi:10.1088/0143-0807/19/3/005.\\
The proof that the authors construct for the limitation on the superposition of state vectors corresponding to different sectors of the Hilbert space, partitioned by a superoperator has a flaw as outlined below. 
\end{abstract}


\section{Introduction}

In Ref.~\cite{ref1}, section 2.4(Impossibility of superposing states belonging to different coherent sectors), the authors construct a proof using a superoperator G that commutes with all observables of the system. The eigenstates of operator G are of the form $|g_m;\alpha_m \rangle$ where $g_m$ and $\alpha_m$ are labels to distinguish different eigenvectors all of which are non-degenerate, i.e. $\langle g_m;\alpha_m | g_n;\alpha_n \rangle = 0$ if $m \neq m$ .\\ In the next section(2.4),
they construct a vector $|u\rangle$ = $\sum_m u_m |g_m;\alpha_m\rangle$ superposing the eigenstates of G and claim that since G commutes with every other observable, $|u\rangle$ should be an eigenstate of every operator in a complete set of commuting
observables of the system.\\
Now we know that if $G|a\rangle = a|a\rangle$ and $G|b\rangle = b|b\rangle$ are eigenvectors of an operator G, then $\lambda_1 |a\rangle+ \lambda_2 |b\rangle$ is not an eigenvector unless they have the same eigenvalue since:
\begin{equation}
G\left(\lambda_1 |a\rangle+ \lambda_2 |b\rangle\right)$ = $ a \lambda_1 |a\rangle+ b \lambda_2 |b\rangle \neq  \lambda \left( \lambda_1 |a\rangle+ \lambda_2 |b\rangle \right)
\end{equation}
(for some $\lambda$) unless a = b, which is not true in general and definitely not true for the hermitian operators with non-degenerate eigenvalues in the paper.

It is true that if G commutes with every other observable and has non-degenerate eigenvalues, then all such obervables share the same eigenvectors. But $|u\rangle$ itself is not an eigenvector of G as proved above. And so $|u\rangle$  cannot be an eigenvector of other observables too. 
To summarize, the proof of superselection rules proposed is thus is not valid for the case discussed by the authors. Whether a similar proof can be constructed is still open.

\end{document}